\documentclass[a4paper,11pt]{article}
\pdfoutput=1 

\usepackage{jcappub} 

\usepackage[T1]{fontenc} 

\title{\boldmath Signatures of Spatial Curvature on Growth of Structures}

\author{Mohammad H. Abbassi}
\author{Amir H. Abbassi}

\affiliation{Department of Physics, School of Sciences, Tarbiat Modares University. P.O. Box 14115-175, Tehran, Iran}

\emailAdd{mh.abbassi@modares.ac.ir}
\emailAdd{ahabbasi@modares.ac.ir}

\abstract{We write down Boltzmann equation for massive particles in a spatially curved FRW universe and solve the approximate line-of-sight solution for evolution of matter density, including the effects of spatial curvature to the first order of approximation. It is shown that memory of early time gravitational potential is affected by presence of spatial curvature. Then we revisit Boltzmann equation for photons in the  general FRW background. Using it, we show that how the frequency of oscillations and damping factor (known as Silk damping) changed in presence of spatial curvature. At last, using this modified damping factor in hydrodynamic regime of cosmological perturbations, we find our analytic solution which shows the effects of spatial curvature on growing mode of matter density.}

\begin{document}
\maketitle
\flushbottom

\section{Introduction}
\label{sec:int}

According to Planck 2018 results \cite{Aghanim:2018eyx} $\Omega_K=-0.044^{+0.018}_{-0.015}$ (68 $\%$ Planck TT, TE, EE + low E) and the joint constraint with lensing and BAO measurements constraint it to $\Omega_K=0.001 \pm 0.002$. The constraint on spatial curvature assumes a specific cosmological model. This means that most of this results are cosmological model dependent. Many model independent methods to measure the spatial curvature of the universe are proposed \cite{Bernstein:2005en,Rasanen:2014mca,Sapone:2014nna,Li:2014yza, Denissenya:2018,Wei:2018,Li:2019,Wei:2020}. These results shows that non-zero $\Omega_K$ can not be easily ruled out by current observations (e.g. look at the results of \cite{Ooba:2017,Handley:2019,Valentino:2019,Valentino:2020}). In addition, there are vast majority of study to constraint topology of the universe using cosmological data \cite{LachiezeRey:1995kj,Cornish:2003db,Bielewicz:2010bh,Aurich:2013fwa}.
Beside this practical concerns, from the theoretical perspective it is also interesting to know how the topology of the universe can affect our cosmological phenomena. Furthermore, data indicates a positive cosmological constant, which leads to a de Sitter universe for vacuum solutions. Knowing that from all the de Sitter solutions only the Lorentzian de Sitter spacetime which is spatially closed, is maximally symmetric, maximally extended and geodesiclally complete, increased our theoretical interests for studying cosmology around this background \cite{Hawking:1973uf,Abbassi}. There is extensive studies on the subject of cosmological perturbations using  Boltzmann hierarchy equations approach in non-flat universe. Study of some of the cosmological observables (like Hubble diagram, angular size and number density of galaxies) in closed universe models can be found in \cite{Bjornsson:1995} and the references therein.   In \cite{Peebles:1970,Wilson:1981,Bond:1984} solving Boltzmann equations is used for calculations on CMB anisotropy. But their approach for spatially curved universe leads to long integration time. The authors of \cite{White:1996} derived CMB angular power spectrum and matter transfer function in a closed FRW universe using a semi-analytic method. In \cite{Allen:1995} the semi-analytic is used to compute CMB tensor spectrum in closed universe. In \cite{Zaldarriaga:1998}, integral solutions are derived for CMB anisotropy in general FRW background. The solutions are in the  form of time integral over source term and geometrical term. The geometrical term is described as eigenfunction of Laplacian operator in spatially curved background. In their paper, they give a fast and efficient method for computing those functions. This approach is generalized in \cite{Hu:1998}  to arbitrary perturbation type and FRW metric. Furthermore, \cite{Lewis:1999} use a numerical approach to linearized equations of $1+3$ covariant formalism to calculate CMB anisotropies  in general FRW background. A new method for calculation of CMB anisotropy in non-flat universe is presented in \cite{Lesgourgues:2013}, in which a new efficient algorithm is introduced to calculate eigenfunctions of Laplacian operators in non-flat universe. Corrections on spatial curvature due to relativistic effects on Large Scale Structure is studied in \cite{Dio:2016}. Some other new investigation of cosmological perturbations and Boltzmann  hierarchies are presented in \cite{Pitrou:2020, Noh:2020}.  It's well-known that spatial curvature causes shifts of angular scale of acoustic peaks, change primordial spectrum and evolution of long wavelength modes through Integrated Sachs-Wolf effect.  In this work we want to explore the effects of spatial curvature in two phenomena  by studying relativistic Boltzmann equation. First it is shown that how spatial curvature affects the memory of early time gravitational potential. The second phenomena that is presented is about damping of acoustic peaks due to transportation of photons in non-relativistic plasma, which is known as silk damping \cite{Silk:1967}. In this regard, we follow closely the procedure of \cite{Weinberg:2008zzc} to drive an analytical formula that shows how frequency and damping of acoustic oscillations are affected by spatial curvature. Especially, we try to give an analytical solution to our equations so that we can have a clear interpretation and perception about behavior of spatial curvature in our solutions.
In section II we write relativistic Boltzmann equation for dark matter for a spatially curved background. We find an integro-differential solution for zeroth moment of it which gives evolution of matter density. In section III by writing relativistic Boltzmann equation for photons we find the effects of curvature on damping parameter and oscillation frequency. In section VI we use the derived damping factor in hydrodynamic limit of cosmological perturbation to show how spatial curvature affects the growth of matter density and especially the ripples of Baryon Acoustic Oscillations. In the last section we summarize our results and give some comments.\\
Throughout this paper we use $-+++$ signatures. Greek alphabets is set for 4 dimensional indices and Latin alphabet for 3 dimensional ones.

\section{Relativistic Boltzmann Equation for Matter}
\label{sec:BoltzMatt}

Through out this paper we assume a background of general curved FRW universe with the metric :
\begin{equation}
\bar{g}_{00}=-1, \quad \bar{g}_{0i}=0, \quad \bar{g}_{ij}=a^2\gamma_{ij}
\end{equation}
where $\gamma_{ij}$ is the 3-dimensional spatial metric. In a semi-Cartesian coordinate it can be written as:
\begin{equation}
\gamma^{ij}=\delta^{ij}-Kx^ix^j
\end{equation} 
where K is the constant of curvature of the spatial metric.
There is a useful relation for spatial metric in semi-Cartesian coordinates which helps us to simplify the following calculations:
\begin{equation}
\gamma^{ii'}\gamma^{jj'}\partial_k\gamma_{i'j'}=-\partial_k\gamma^{ij}
\end{equation}
Furthermore, the non vanishing components of the Christoffel's connection of the metric are:
\begin{align}\label{christ}
\bar{\Gamma}_{ij}^k &= \frac{1}{2}\gamma^{kl}\left(\partial_j \gamma_{li}+\partial_i\gamma_{lj}-\partial_l\gamma_{ij}\right)\\
\bar{\Gamma}_{0i}^j &=H\delta^j_i\\
\bar{\Gamma}^0_{ij} &=a^2 H \gamma_{ij}
\end{align}

We define the four-momentum of dark matter particles as:
\begin{equation}
p^\mu=m \frac{dx^\mu}{d\tau}
\end{equation}
In this way, the particle's four-momentum is related to particle's velocity by $\frac{dx^i}{dt}=\frac{p^i}{p^0}$ and zeroth component of four momentum would be $p^0=\sqrt{m^2+g_{ij}p^ip^j}$. To write down the Boltzmann equation, we should define the number density of the particles as function of Cartesian coordinates $x^i$, momenta $p_i$ and time $t$ which are the usual phase space parameters of a system of particles. Note that here we use lower index $p_i$ only because the later calculations are less complicated in terms of them . With the help of geodesic equation, we get:
\begin{equation}\label{geoeq}
\frac{dp_i}{dt}=\partial_ig_{mn}\frac{p^mp^n}{2p^0}
\end{equation}
Because the interactions of the cold dark matter particles are only restricted to the gravitational interaction, number density of cold dark matter satisfies collisionless Boltzmann equation, which is nothing but the fact that number density is conserved in phase space \cite{Dodelson:2020}. After applying chain rules and with the help of (\ref{geoeq}) we will have:
\begin{equation}\label{boltzeqn}
\frac{d}{dt}n(x^i,p_i,t)=\partial_t n(x^i,p_i,t) +\frac{p^k}{p^0}\partial_k n(x^i,p_i,t)+\frac{p^lp^m}{2p^0}\partial_k g_{lm}\frac{\partial n(x^i,p_i,t)}{\partial p_k}=0
\end{equation}
Then, we write the metric perturbations around this background as:
\begin{equation}\label{pert}
g_{ij}=a^2\gamma_{ij}+\delta g_{ij}, \quad g^{ij}=a^{-2}\gamma^{ij}-a^{-4}\delta g^{ij}
\end{equation}
where the covariant and contravariant forms of the metric perturbations in the above definitions, are related in this way:
\begin{equation}
\delta g^{ij}=\gamma^{ii'}\gamma^{jj'}\delta g_{i'j'}
\end{equation}
After linearizing the zeroth and i-th components of 4-momenta of the particles with respect to the metric perturbation we get:
\begin{equation}\label{p0}
p^0=\sqrt{m^2+p^2/a^2}\left(1-\frac{a^{-4}p_ip_j\delta g^{ij}}{2(m^2+p^2/a^2)}\right)
\end{equation}
\begin{equation}
p^i=a^{-2}\gamma^{ij}p_j-a^{-4}p_j\delta g^{ij}
\end{equation}

in which p is defined as $p\equiv\sqrt{\gamma^{ij}p_ip_j}$. The number density of the particles can be written as its background value plus perturbed number density where the background number density is function of $a\sqrt{g^{ij}p_ip_j}$: 
\begin{equation}
n=\bar{n}\left(a\sqrt{g^{ij}p_ip_j}\right)+\delta n(x^i,p_i,t)
\end{equation} 
The factor $a(t)$ is included in its argument so that  $\bar{n}(p)$ becomes time independent. After linearizing the background number density in terms of metric perturbations, we get:
\begin{equation}\label{nbar}
\bar{n}=\bar{n}\left(a\sqrt{g^{ij}p_ip_j}\right)=\bar{n}(p)-\bar{n}'(p) \frac{p_ip_j\delta g^{ij}}{2 a^2 p}
\end{equation}

After straightforward but tedious calculations it can be shown that:
\begin{equation}\label{bgn}
\frac{p^k}{p^0}\partial_k \bar{n}+\frac{p^lp^m}{2p^0}\partial_k g_{lm}\frac{\partial\bar{n}}{\partial p_k}=0
\end{equation}

Putting the above number density in Boltzmann equation (\ref{boltzeqn}) and using (\ref{bgn}), we would have our Boltzmann equation for perturbed number density:
\begin{equation}\label{Boltz1}
\partial_t\delta n  +\frac{a^{-2} \gamma^{ij}p_j}{\sqrt{m^2+p^2/a^2}}\frac{\partial\delta n}{\partial x^i}+\frac{a^{-2}K \vec{x}.\vec{p}p_k}{\sqrt{m^2+p^2/a^2}}\frac{\partial \delta n}{\partial p_k}=n'(p)\frac{p_ip_j\partial_t\left(a^{-2}\delta g^{ij}\right)}{2p}
\end{equation}
The Fourier expansion of the scalar modes in a spatially closed universe can be written in the following form:
\begin{equation}\label{expan}
\delta n(x^i,p_i,t)=\int d^2\hat{q}\sum_q \frac{e^{iq \arccos(\hat{q}.\vec{x})}}{\sqrt{1-(\hat{q}.\vec{x})^2}}\delta n(q,\hat{q}_i,p_i,t)
\end{equation}
where  q is non-negative integer,  which is the well-known fact that wave numbers are discrete in spatially closed universe. In the same way, we can expand the metric perturbation, $\delta g_{ij}$.
Putting these modes expansion back in  Boltzmann equation (\ref{Boltz1}) and working in limit of $x\ll 1$, the equation (\ref{Boltz1}) takes this form,
\begin{equation}\label{Bolts2}
\partial_t\delta n(q_i,p_i,t)+ \frac{a^{-2}}{\sqrt{m^2+p^2/a^2}} \mathcal{O} \delta n(q_i,p_i,t)= \frac{\bar{n}'(p)}{2p}p_ip_j\partial_t\left(a^{-2}\delta g^{ij}(q_i,t)\right)
\end{equation}
where $q_i=q\hat{q}_i$ and we define the differential operator $\mathcal{O}$ as:
\begin{equation}
\mathcal{O}\equiv i\vec{q}.\vec{p}+iKp_kp_l \frac{\partial^2}{\partial p_k\partial q_l}
\end{equation}
From now till the end of this section, we omit the arguments of the perturbation parameters, so that are relations looks much simpler. But we should note that these perturbations are in Fourier space.  The integral solution of above differential equation has the following form: 
\begin{equation}
\delta n =\int_{t_1}^t dt' exp\left[-\int_{t'}^t dt'' \frac{1}{a^2(t'')\sqrt{m^2+p^2/a^2(t'')}}\mathcal{O} \right]\frac{\bar{n}'(p)}{2p}p_ip_j\partial_t\left(a^{-2}\delta g^{ij}\right)
\end{equation}

By taking different moments of number density, we can get  each components of energy-momentum tensor. Here we we only use its zeroth moment which is :
\begin{equation}\label{T00}
T^0_0=\frac{1}{\sqrt{-g}}\int d^3p \left(p_0 n \right)
\end{equation}

At the end,  the fluid parameters can be extracted from different components of energy-momentum tensor, $\delta T^0_0=-\delta \rho$.\\

In this section, we want to choose the metric perturbations in Newtonian gauge. Because the form of our solutions will be simpler in this gauge. From Scalar-Vector-Tensor decomposition, we know that our perturbed equations do not mix different types of perturbations. Because our goal is to derive matter density, it is just enough to focus on scalar part of the metric perturbations:
\begin{equation}
ds^2=-(1+2\Phi)dt^2+a^2(1-2\Psi)\gamma_{ij} dx^idx^j
\end{equation}
From ij component of Einstein perturbed equation, it is easy to see that in the absence of scalar stress tensor, $ \Pi^s$, the Newtonian potentials $\Phi$  and $\Psi$ are equal.\\
Here, the metric perturbation, $\Phi$, behaves as a source on the right hand side of equation (\ref{Bolts2}). To first order in metric perturbation we would have, $\sqrt{-det(g)}= \frac{a^3}{\sqrt{1-Kx^2}} (1-2\Phi)\approx a^3(1-2\Phi)$ . Also the parameters  $p^0$ and $\bar{n}$ are perturbed as (\ref{p0}) and  (\ref{nbar}).
Inserting the integral solution of number density in (\ref{T00}) we will find:
\begin{equation}
\begin{split}
\delta \rho&=\frac{1}{a^3}\int d^3p( -\frac{\bar{n}(p)\Phi}{\sqrt{m^2+p^2/a^2}}(2m^2+3p^2/a^2)-\bar{n}'(p)\Phi p\sqrt{m^2+p^2/a^2}\\
& +\sqrt{m^2+p^2/a^2}\int_{t_1}^t dt' exp\left[-\int_{t'}^t dt'' \frac{1}{a^2(t'')\sqrt{m^2+p^2/a^2(t'')}}\mathcal{O} \right] \bar{n}'(p)p\partial_t\Phi)
\end{split}
\end{equation}
After using by part in the second integral, it will be simplified like this,
\begin{footnotesize}
\begin{equation}\label{T001}
\begin{split}
\delta \rho= & \frac{1}{a^3}\int d^3p \sqrt{m^2+p^2/a^2}\left(-\bar{n}(p)\Phi+\int_{t_1}^t dt' exp\left[-\int_{t'}^t dt'' \frac{1}{a^2(t'')\sqrt{m^2+p^2/a^2(t'')}}\mathcal{O} \right] \bar{n}'(p)p\partial_t\Phi \right)
\end{split}
\end{equation}
\end{footnotesize}
Using (\ref{T00}) and after linearizing the (\ref{T001}) in terms of K, $\delta \rho$ can be written as:
\begin{equation}
\begin{split}
\delta \rho = -\bar{\rho}\Phi + &\frac{1}{a^3}\int d^3p \sqrt{m^2+p^2/a^2}\int_{t_1}^t dt' exp\left[-\int_{t'}^t dt'' \frac{i\vec{q}.\vec{p}}{a^2(t'')\sqrt{m^2+p^2/a^2(t'')}}\right]\\
&\left( \bar{n}'(p)p\partial_{t'}\Phi-iK\frac{\vec{p}.\vec{q}}{q}(p^2\bar{n}''(p)+p\bar{n}'(p))\partial_{t'}\Phi' \int_{t'}^t\frac{dt''}{a(t'')^2\sqrt{m^2+p^2/a(t'')^2}}\right)
\end{split}
\end{equation}

Here the prime over $\Phi$ indicates derivative with respect to wave number q. The momentum space volume element in the limit of small scales can be written as,  $d^3p=p^2 dpdzd\phi$, where $z=\frac{\vec{p}.\vec{q}}{qp}$, is the cosine of angle between $\vec{p}$ and $\vec{q}$, $\phi$ is azimuthal angle of $\vec{p}$ with respect to $\vec{q}$.
So we can re-parametrize the integral solution in terms of these new parameters:
\begin{equation}
\begin{split}
\delta \rho= -\bar{\rho}\Phi+\frac{2\pi}{a^3}\int_0^\infty p^2dp&\int_{-1}^1dz\sqrt{m^2+p^2/a^2}\int_{t_1}^t dt' exp\left[-\int_{t'}^t dt''\frac{iqpz}{a(t'')^2\sqrt{p^2/a(t'')^2+m^2}}\right]\\
&\left(\bar{n}'(p)p\partial_{t'}\Phi-\int_{t'}^t dt''\frac{iKzp^2}{a(t'')^2\sqrt{p^2/a(t'')^2+m^2}}(p\bar{n}''(p)+\bar{n}'(p))\partial_{t'}\Phi'    \right)
\end{split}
\end{equation}

Doing this integral solution in general is very complicated, but it will have much more simpler form in non relativistic limit where $p^2/a^2\ll m^2$, which is a permissible assumption for cold dark matter particles:
\begin{equation}
\begin{split}
\delta \rho=-\bar{\rho}\Phi +\frac{2\pi}{a^3}&\int dp p^2\int dz\int_{t_1}^t dt' exp\left[-\frac{iqpz}{m}\int_{t'}^t\frac{dt''}{a^2(t'')}\right]\\
& \left(m\bar{n}'(p)p\partial_{t'}\Phi -iKzp^2(p\bar{n}''(p)+\bar{n}'(p))\partial_{t'}\Phi'\int_{t'}^t \frac{dt''}{a(t'')^2} \right)
\end{split}
\end{equation}

Also we should note that in the non relativistic limit, the rate of oscillation in the above exponential is much smaller than the rate of oscillation in metric perturbation, $\Phi$ \cite{Flauger:2017ged}. So we can pull back the time derivative to the whole of the argument:
\begin{equation}
\begin{split}
exp&\left[-\frac{iqpz}{m}\int_{t'}^t\frac{dt''}{a^2(t'')}\right] \left(m\bar{n}'(p)p\partial_{t'}\Phi -iKzp^2(p\bar{n}''(p)+\bar{n}'(p))\partial_{t'}\Phi'\int_{t'}^t \frac{dt''}{a(t'')^2} \right) \\
&\approx  \partial_{t'}\left(exp\left[-\frac{iqpz}{m}\int_{t'}^t\frac{dt''}{a^2(t'')}\right] \left(m\bar{n}'(p)p\Phi -iKzp^2(p\bar{n}''(p)+\bar{n}'(p))\Phi'\int_{t'}^t \frac{dt''}{a(t'')^2} \right)\right)
\end{split}
\end{equation}  
Using this fact, we can simplify $\delta \rho$ in this manner:
\begin{footnotesize}
\begin{equation}
\delta \rho=-\bar{\rho}\Phi+\frac{2\pi}{a^3}\int dp p^2 dz \biggl( m\bar{n}'(p)p\Phi(t)-e^{-\frac{iqpz}{m}\int_{t_1}^t\frac{dt''}{a^2(t'')}}(m\bar{n}'(p)p\Phi(t_1)-iKzp^2(n''(p)p
              +n'(p))\Phi'(t_1)\int_{t_1}^t\frac{dt''}{a^2(t'')} ) \biggr)
\end{equation}
\end{footnotesize}
with the help of by part in the first term of the above integral and using the fact that  $\frac{m \mathcal{N}}{a^3}\equiv \frac{m}{a^3}\int 4\pi p^2dp \bar{n}=\bar{\rho}$ and after integration over z, we get:

\begin{equation}\label{eq37}
\delta\equiv \frac{\delta \rho}{\bar{\rho}}=-2\Phi +\mathcal{S}
\end{equation}
\begin{footnotesize}
\begin{equation}\label{ddd}
\mathcal{S}\equiv \frac{m\Phi(t_1)}{\mathcal{N}\omega_q(t)}\int p^2 dp \bar{n}'(p)Sin(\frac{p\omega_q(t)}{m})\\
 -\frac{K\Phi'(t_1)}{\mathcal{N}q}\int p^3 dp (\bar{n}''(p)p+\bar{n}'(p))\left(Cos(\frac{p\omega_q(t)}{m})-\frac{Sin(\frac{p\omega_q(t)}{m})}{\frac{p\omega_q(t)}{m}}\right)
\end{equation}
\end{footnotesize}
where the parameter $\omega_q(t)$ is defined as:
 \begin{equation} 
 \omega_q(t)\equiv \int_{t_1}^t \frac{qdt''}{a^2(t'')} 
 \end{equation}
 In the absence of spatial curvature, the second term can be interpreted as memory of the gravitational field at early times. An effect like this find in \cite{Flauger:2017ged} for gravitational waves, where they show that the presence of dark matter affects propagation of gravitational wave by its memory at the time of emission. Here it can be seen that the spatial curvature modified this memory as the second term of (\ref{ddd}). The function $\mathcal{S}(t)$ is a transient function that goes to zero exponentially  at late time, when dark matter particles travel a distance larger than mode's wavelength, or in other words when $\frac{p\omega_q(t)}{m} \gg 1$. To clarify this fact, let us assume that $n(p)$ has familiar Maxwell-Boltzmann form of:
\begin{equation}
\bar{n}(p)=A e^{-\frac{p^2}{2P^2}}
\end{equation}
and we set the normalization constant to $A=\sqrt{\frac{2}{\pi}}\frac{\mathcal{N}}{4\pi P^3}$ so that $\int d^3p\bar{n}=\mathcal{N}$ , where $P$ is an arbitrary constant.

Then after performing by parts and doing integration over p we will end up with:
\begin{equation}\label{end1}
\mathcal{S} = \left( \Phi(t_1)(\bar{v^2}\omega_q(t)^2-3)+\frac{K\Phi'(t_1)}{q}\omega_q(t)^2\bar{v^2}(25-13\omega_q(t)^2\bar{v^2}+\omega_q(t)^4\bar{v^2}^2)\right)e^{-\frac{\omega_q(t)^2\bar{v^2}}{2}}
\end{equation}
We define $\bar{v^2}=\frac{P^2}{m^2}$, which is the mean square coordinate velocity for distribution function of $\bar{n}$.  As we mentioned earlier, dark matter particles are non relativistic and so  $\bar{v^2}/a^2(t) \ll 1 $. For the particles that travel less than wavelength of the modes between $t_1$ and t, we can assume  $\bar{v^2}\omega_q(t)^2 \ll 1$. Then we have $\mathcal{S}=-3\Phi(t_1)$.
In addition, at very late time, where $\bar{v^2}\omega_q^2 \gg 1$, $\mathcal{S}(t)$ is exponentially small and  this memory effect of the early time  will be erased. In the between time, when $ v^2\omega_q^2\sim 1$, we can see that spatial curvature, shows itself in memory as second term of (\ref{end1}).

\section{Damping of Photons due to Non relativistic Medium}
\label{sec:SilkDamping}
The goal of this part is to compute the oscillation frequency and damping factor of acoustic peaks in curved FRW background. Here we closely follow the procedure of \cite{Weinberg:2008zzc} in deriving our results.\\ 
In the case of photons, we denote their distribution function by $n^{ij}$ and the upper indices are due to photon's polarization. This function obeys the Boltzamann equation as:
\begin{equation}
\begin{split}
\partial_t n^{ij}+\frac{p^k}{p^0}\partial_k n^{ij}+\frac{p^lp^m}{2p^0}\partial_k g_{lm}\frac{\partial n^{ij}}{\partial p_k}+
(\Gamma^i_{k\lambda}-\frac{p^i}{p^0}\Gamma^0_{k\lambda})\frac{p^\lambda}{p^0}n^{kj}+(\Gamma^j_{k\lambda}-\frac{p^j}{p^0}\Gamma^0_{k\lambda})\frac{p^\lambda}{p^0}n^{ki}=C^{ij}
\end{split}
\end{equation}
Here $C^{ij}$ is the collision term, which includes interactions of Thomson scattering of photons with electrons in the plasma. Then, we can write the perturbed distribution function of photon as:

\begin{equation}
n^{ij}=\frac{1}{2}\bar{n}(ap^0)(g^{ij}-\frac{g^{ik}g^{jl}p_kp_l}{(p^0)^2 })+\delta n^{ij}
\end{equation} 
We can follow the same procedure of the last section, now for photon distribution. Considering the metric perturbation (\ref{pert}), the perturbed distribution function becomes: $\bar{n}=\bar{n}(p)-\frac{p_ip_j}{2a^2p}\bar{n}'(p)\delta g^{ij}$. In addition, it will be easy to show that $\bar{n}$ satisfies the same relation as (\ref{bgn}). Using this and Christoffel values of (\ref{christ}) streaming part of Boltzmann equation would be:
\begin{equation}
\frac{d}{dt}n^{ij}=\partial_t \delta n^{ij}+\frac{1}{a}\gamma^{kl}\hat{p}_l\partial_k \delta n^{ij}+\frac{2K}{a}\hat{p}_kp_lx^l\frac{\partial \delta n^{ij}}{\partial p_k}-\frac{p}{4a^2}\hat{p}_{i'}\hat{p}_{j'}\bar{n}'\partial_t(a^{-2}\delta g^{i'j'})(\gamma^{ij}-\gamma^{ik}\gamma^{jl}\hat{p}_k\hat{p}_l)\\
+2\frac{\dot{a}}{a}\delta n^{ij}
\end{equation}
The $\hat{p}_i$ is the unit vector of $p_i$, which is defined as $\hat{p}_i\equiv\frac{p_i}{p}$. Furthermore, the collision part can be written as:
\begin{equation}
\begin{split}
C^{ij}=&-\omega_c\delta n^{ij}+\frac{3\omega_c}{8\pi}\int d\hat{p}_1[\delta n^{ij}(x^i,p\hat{p}_1,t)-\gamma^{ik}\hat{p}_k\hat{p}_l\delta n^{lj}(x^i,p\hat{p}_1,t)-\gamma^{jk}\hat{p}_k\hat{p}_l\delta n^{il}(x^i,p\hat{p}_1,t)\\
&+\gamma^{ik}\gamma^{jl}\hat{p}_k\hat{p}_l\hat{p}_m\hat{p}_n \delta n^{mn}(x^i,p\hat{p}_1,t)]
-\frac{\omega_c}{2a^3}p_k\delta u_{k'}\gamma^{kk'}\bar{n}'[\gamma^{ij}-\gamma^{il}\gamma^{jl'}\hat{p}_l\hat{p}_{l'}]
\end{split}
\end{equation}
where the definition of $p_1$ in above relation is :
\begin{equation}
p_1=p(1+\frac{(\hat{p}_1-\hat{p})_k\delta u_{k'}\gamma^{kk'}}{a})
\end{equation}
Our goal is to derive damping of acoustic waves in a homogeneous and time-independent plasma in a spatially curved background.  So we can consider a static universe $a=1$ with spatial curvature as:
\begin{equation}
ds^2=-dt^2+\gamma_{ij}dx^idx^j
\end{equation}
When the collision rate is much larger than sound frequency, we can also assume that $\omega_c$ is time independent.
This assumption of static universe is permissible, because the damping effect would only be noticeable for deep inside the horizon modes. So we can put our Boltzmann equation in this form:
\begin{footnotesize}
\begin{equation}\label{boltz}
\begin{split}
&\partial_t \delta n^{ij}+\gamma^{kl}\hat{p}_l\partial_k \delta n^{ij}+2K\hat{p}_k p_lx^l\frac{\partial \delta n^{ij}}{\partial p_k}=-\omega_c\delta n^{ij}+\frac{3\omega_c}{8\pi}\int d\hat{p}_1[\delta n^{ij}(x^i,p\hat{p}_1,t)-\gamma^{jk}\hat{p}_k\hat{p}_k\hat{p}_l\delta n^{il}(x^i,p\hat{p}_1,t)\\
&+\gamma^{ik}\gamma^{jl}\hat{p}_k\hat{p}_l\hat{p}_m\hat{p}_n\delta n^{mn}(x^i,p\hat{p}_1,t)]-\frac{\omega_c}{2}p_k\delta u_{k'}\gamma^{kk'}\bar{n}'(\gamma^{ij}-\gamma^{il}\gamma^{jl'}\hat{p}_l\hat{p}_{l'})
\end{split}
\end{equation}
\end{footnotesize}
Then we can expand $\delta n^{ij}(t,\vec{x},\vec{p})$ and $\delta u_j(t,\vec{x})$ like the mode expansions of (\ref{expan}). Because we want to consider the acoustic modes deep inside the horizon, we can use the same approximation of the last section, $x\ll 1$.
After using this expansion in equation (\ref{boltz}), we get:
\begin{footnotesize}
\begin{equation}\label{boltz1}
\begin{split}
\int d^3p(\omega_c-i\omega+i\vec{q}.\hat{p})\delta n^{ij}&+2iK\hat{p}_kp_l\frac{\partial}{\partial p_k}\frac{\partial}{\partial q_l}\delta n^{ij}=-\frac{\omega_c}{2}\bar{n}'p.\delta u(\gamma^{ij}-\hat{p}^i\hat{p}^j)\\
&+\frac{3\omega_c}{8\pi}\int d\hat{p}_1 (\delta n^{ij}(p\hat{p}_1)-\hat{p}^i\hat{p}_l\delta n^{lj}(p\hat{p}_1)
-p^j\hat{p}_l\delta n^{il}(p\hat{p}_1)+\hat{p}^i\hat{p}^j\hat{p}_m\hat{p}_n\delta n^{mn}(p\hat{p}_1))
\end{split}
\end{equation}
\end{footnotesize}
Then, we multiply $p_j$ with $kk$ component of equation (\ref{boltz1}), so that we come to:

\begin{equation}\label{pkk}
\begin{split}
&\int d^3p (\omega_c-i\omega+i\vec{q}.\hat{p})p_j\delta n^{kk}+2iK\int d^3p \hat{p}_lp_{l'}p_j\frac{\partial}{\partial p_l}\frac{\partial}{\partial q_{l'}}\delta n^{kk}=-\omega_c\int d^3 p p_j p_k \delta u_k \bar{n}'
\end{split}
\end{equation}
Using by part for derivative of $p_l$, the second term of left hand side of equation (\ref{pkk}) gives:
\begin{equation}\label{eq1a}
\begin{split}
&2iK\int d^3p \hat{p}_lp_{l'}p_j\frac{\partial}{\partial p_l}\frac{\partial}{\partial q_{l'}}\delta n^{kk}=-2iK\int d^3p(3\hat{p}_{l'}p_j+\hat{p}_{l'}p_j+\hat{p}_{l'}p_j)\frac{\delta n^{kk}}{\partial q_{l'}}=-10 iK\int d^3p p_j \hat{p}_{l'}\frac{\delta n^{kk}}{\partial q_{l'}}
\end{split}
\end{equation}
For simplifying the term on the right hand side of (\ref{pkk}), we can write $\int d^3 p p_jp_k\bar{n}'(p)=H_{ij}$. Contracting it with $\gamma_{ij}$ gives :
\begin{equation}\label{eq1b}
\gamma^{ij} H_{ij}=\int d^3 pp^2\bar{n}'(p)=-4\int d^3p p \bar{n}(p)=-4\bar{\rho}_{\gamma}
\end{equation}
In which $\bar{\rho}_\gamma$ is photon energy density. So we have: $H_{ij}=-\frac{4}{3}\bar{\rho}_{\gamma}\gamma_{ij}$.
Using (\ref{eq1a}) and knowing $H_{ij}$ we can rewrite equation (\ref{pkk}):
\begin{equation}\label{theeq}
\begin{split}
&\int d^3p(\omega_c-i\omega+i\vec{q}.\hat{p})p_j\delta n^{kk}=10iK\int d^3p p_j\hat{p}_l \frac{\partial}{\partial q_l}\delta n^{kk}+\frac{4}{3}\omega_c\bar{\rho}_{\gamma}\delta u_j
\end{split}
\end{equation}
The stress tensor for photon distribution function is defined like this:
\begin{equation}
{\delta T^i_j}_\gamma=\gamma^{ii'}\int d^3p n^{kk}p\hat{p}_{i'}\hat{p}_j, \quad {\delta T^0_j}_\gamma=\int d^3p n^{kk}p\hat{p}_j
\end{equation}
For Baryon we have: $\delta {T_B}^i_j=0$ and $ \delta {T_B}^0_j=\bar{\rho}_B\delta u_j$.
Using momentum conservation, $\delta T^0_j + \delta T^i_j=0$, we can write:
\begin{equation}\label{eqcons}
\begin{split}
&\omega\bar{\rho}_B\delta u_j=-\int d^3p n^{kk}(p)\hat{p}_j(\hat{p}_i q_i-\omega)
\end{split}
\end{equation}
With the help of (\ref{eqcons}), the equation (\ref{theeq}) simplifies to:
\begin{equation}\label{H43}
(\omega \rho_B+i\frac{4}{3}\omega_c\rho_{\gamma})\delta u_j=i\int d^3p p_j(\omega_c\delta n^{kk}-i10K\hat{p}_l\frac{\partial}{\partial q_l}\delta n^{kk})\\
\end{equation}
Now we can define form factors in this way:
\begin{equation}
\int d^3p p \delta n^{ij}=\bar{\rho}_{\gamma} (X\delta_{ij}+Y\hat{q}_i\hat{q}_j)
\end{equation}
\begin{equation}\label{pjnkk}
\int d^3 p p_j \delta n^{kk} =\bar{\rho}_{\gamma}Z\hat{q}_j
\end{equation}
\begin{equation}
\int d^3 p \delta n^{kk} p_jp_j=\bar{\rho}_{\gamma}(V\delta_{ij}+W \hat{q}_i\hat{q}_j)
\end{equation}
These integrals can be written in this way, because after integrating over $\vec{p}$, the only parameter that contains direction is $q_i$. Using these form factors in equation (\ref{H43}) leads to:
\begin{equation}
\begin{split}
&(\omega \bar{\rho}_B+i\frac{4}{3}\omega_c \bar{\rho}_{\gamma})\delta u_j=i\omega_c\bar{\rho}_{\gamma}\hat{q}_j Z+\frac{10K\bar{\rho}_{\gamma}}{q}\frac{\partial}{\partial \hat{q}_l}(V\delta _{lj}+W \hat{q}_l\hat{q}_j)=i\omega_c\bar{\rho}_{\gamma}Z\hat{q}_j+4W\frac{\hat{q}_j}{q}
\end{split}
\end{equation}
\begin{equation}\label{equ}
(1-i\omega R t_c)\delta u_j=\frac{3}{4}\hat{q}_j(Z-i\frac{40Kt_c}{q}W)
\end{equation}
where $R=\frac{3\bar{\rho}_\gamma}{4\bar{\rho}_B}$. From the relation (\ref{equ}), the perturbed velocity can be written: $\delta u_j = \frac{3\hat{q}_j}{4(1-i\omega t_c R)}(Z-i\frac{40Kt_c}{q}W)$.
Putting this back to equation (\ref{boltz1}), gives:
\begin{footnotesize}
\begin{equation}\label{EQ}
\begin{split}
(\omega_c-i\omega+i\vec{q}.\hat{p})\delta n^{ij}+2iK\hat{p}_l\hat{p}_{l'}\frac{\partial}{\partial p_l}\frac{\partial}{\partial \hat{q}_{l'}}\delta n^{ij}&=\frac{3\omega_c}{8\pi}\int d^2\hat{p}_1(\delta n^{ij}(p\hat{p}_1)-\hat{p}^i\hat{p}_l\delta n^{lj}-\hat{p}^j\hat{p}_l\delta n^{il}+\hat{p}^i\hat{p}^j\hat{p}_m\hat{p}_n\delta n^{mn})\\
&-\frac{\omega_c}{2}\bar{n}'(\delta_{ij}-\hat{p}_i\hat{p}_j)\frac{3\hat{q}_k\hat{p}_k}{4(1-i\omega t_c R)}\left(Z-\frac{i40Kt_c}{q}W\right)
\end{split}
\end{equation}
\end{footnotesize}
We can now integrate right hand side of equation (\ref{EQ}) over $dp p^3$ :
\begin{equation}
\begin{split}
&\int dp p^3 RHS=\frac{3\omega_c}{8\pi}((\delta_{ij}-\hat{p}_i\hat{p}_j)\left(X+\frac{Z-\frac{i40Kt_c}{q}}{1-i\omega t_c R}\hat{q}_k \hat{p}_k\right)+Y(\hat{q}^i-\hat{p}^i\hat{p}.\hat{q})(\hat{q}^j-\hat{p}^j(\hat{p}.\hat{q})))
\end{split}
\end{equation}
After using $\frac{\partial}{\partial p_k}=\hat{p}_k \frac{\partial}{\partial p}$, on the left hand side of (\ref{EQ}), our Boltzmann equation becomes:
\begin{footnotesize}
\begin{equation}\label{beq}
\begin{split}
&(\omega_c-i\omega+i\vec{q}.\hat{p}-8iK\hat{p}_l\frac{\partial}{\partial q_l})\int dp p^3 \delta n^{ij}=\frac{3\omega_c}{8\pi}((\delta_{ij}-\hat{p}_i\hat{p}_j)(X+\frac{4}{3}\delta u_k \hat{p}_k)+Y(\hat{q}^i-\hat{p}^i(\hat{p}.\hat{q}))(\hat{q}^j-\hat{p}^j(\hat{p}.\hat{q}))
\end{split}
\end{equation}
\end{footnotesize}
We define the differential operator $\mathcal{Q}$ like this:
\begin{equation}
\mathcal{Q}\equiv \omega-\vec{q}.\hat{p}+8K\hat{p}_l\frac{\partial}{\partial q_l}
\end{equation} 
Using this definition, the Boltzmann equation (\ref{beq}) gives us:
\begin{equation}
\begin{split}
&4\pi \int dp p^3 \delta n^{ij}=\frac{3\bar{\rho}}{2(1-i\mathcal{Q}t_c)}((\delta_{ij}-\hat{p}_i\hat{p}_j)(X+\frac{Z-i40Kt_cW/q}{1-i\omega t_c R}\hat{p}.\hat{q})+Y(\hat{q}^i-\hat{p}^i(\hat{p}.\hat{q}))(\hat{q}^j-\hat{p}^j(\hat{p}.\hat{q})))
\end{split}
\end{equation}
Then, expanding this equation to second order in $t_c$ gives:
\begin{footnotesize}
\begin{equation}
\begin{split}
4\pi \int dp p^3 \delta n^{ij}=\frac{3}{2}\bar{\rho}_{\gamma}\left(1+i\mathcal{Q}t_c-t_c^2\mathcal{Q}^2\right)\biggl((\delta_{ij}-\hat{p}_i\hat{p}_j)\Bigl(X+(1+i\omega t_c R-t_c^2\omega^2R^2)&(Z-\frac{i40Kt_c}{q}W)\hat{p}.\hat{q}\Bigr)\\
&+Y(\hat{q}^i-\hat{p}^i\hat{p}.\hat{q})(\hat{q}_j-\hat{p}_j\hat{p}.\hat{q})\biggr)
\end{split}
\end{equation}
\end{footnotesize} 
We start by an ansatz that $X$ and $Z$ are of order $O(t_c^0)$ and $Y$ and $W$ are of order $O(t_c)$. At the end we can check that our ansatz was correct. So to second order of $t_c$, we have:
\begin{equation}\label{boltint}
\begin{split}
4\pi \int& dp p^3 \delta n^{ij}=\frac{3}{2}\bar{\rho}_{\gamma}\biggl((\delta_{ij}-\hat{p}_i\hat{p}_j)\Bigl(X+it_c\mathcal{Q}X-t_c^2\mathcal{Q}^2X+((1+it_c\omega R-t_c^2\omega^2 R^2)Z\\
&-\frac{40iK}{q}t_cW)\hat{p}.\hat{q}+it_cZ(1+it_c\omega R)\mathcal{Q}\hat{p}.\hat{q}-t_c^2Z\mathcal{Q}^2\hat{p}.\hat{q}\Bigr)+(1+it_c\mathcal{Q})Y(\hat{q}^i-\hat{q}^i\hat{p}.\hat{q})(\hat{q}^j-\hat{p}^j\hat{p}.\hat{q})\biggr)
\end{split}
\end{equation}
Now we should integrate them over $d^2 \hat{p}$. The calculation for each term of integration is written in the appendix. The final result will be:
\begin{equation}
\begin{split}
4\pi\int d^3p p \delta n^{ij}=&4\pi\rho_{\gamma}(X\delta_{ij}+Y\hat{q}_i\hat{q}_j)=4\pi\rho_{\gamma}\biggl(\Bigl(X+it_cX\omega-\frac{t_c^2}{4}(2\omega^2-16K+\frac{q^2}{5})X+\\
&\frac{8K}{q}(it_c-t_c^2\omega(1+R))Z+(-\frac{2it_cq}{5}+\frac{2t_c^2\omega (R+2)q}{5})Z+\frac{1+i\omega t_c}{10}Y\Bigr)\delta_{ij} \\
&+\Bigl(-\frac{1}{10}t_c^2X+\frac{it_cq}{5}Z-\frac{3\omega RZqt_c^2}{5}+\frac{7(1+i\omega t_c)}{10}Y\Bigr)\hat{q}_i\hat{q}_j\biggl)
\end{split}
\end{equation}
Equating the coefficients of $\gamma_{ij}$ and $\hat{q}_i\hat{q}_j$ on each side correspondingly, will lead to:
\begin{equation}
X(it_c\omega-\omega^2t_c^2+8Kt_c^2-\frac{2}{5}q^2t_c^2)+\frac{1+i\omega t_c}{10}Y+(\frac{2q}{5}(-it_c+t_c^2\omega R+2\omega t_c^2)+\frac{8K}{q}(-t_c^2\omega+it_c-t_c^2\omega R))Z=0
\end{equation}

\begin{equation}
\frac{t_c^2q^2}{5}X+(-\frac{3}{10}+\frac{7i}{10}\omega t_c)Y+(\frac{it_cq}{5}-\frac{2+R}{5}\omega qt_c^2)Z=0
\end{equation}

In addition, using equation (\ref{pjnkk}) and after expanding it to second order in power of $t_c$ we get:
\begin{equation}\label{forapp}
\begin{split}
&4\pi\int d^3p \delta n^{ii}p_j=3\bar{\rho}\int d^2\hat{p} \hat{p}_j\Bigl(X+\frac{Y}{2}+(Z+it_c\omega R Z(1+it_c\omega R)-\frac{it_c40KW}{q})\hat{p}.\hat{q}\\
&+it_c\mathcal{Q}(X+\frac{Y}{2}it_c\mathcal{Q}(Z-\frac{Y}{2}+it_c\omega RZ)\hat{p}.\hat{q}-t_c^2\mathcal{Q}^2X-t_c^2\mathcal{Q}^2Z\hat{p}.\hat{q}-\frac{Y}{2}\hat{p}.\hat{q}^2\Bigr)
\end{split}
\end{equation}
After calculating the integrals of each term (which is written the appendix), we will get:
\begin{footnotesize}
\begin{equation}
\begin{split}
(-it_cq+2\omega t_c^2q)X-\frac{iqt_cq}{5}Y+(it_c\omega(1+R) -t_c^2\omega^2(1+R +R^2)-\frac{3}{5}q^2t_c^2+24Kt_c^2)Z-\frac{i40Kt_cW}{q}=0
\end{split}
\end{equation}
\end{footnotesize}
\\
At last, we can put the the system of equations derived for form factors into a matrix:

\begin{footnotesize}
\begin{equation}
\begin{bmatrix}
it_c\omega-\omega^2t_c^2+8Kt_c^2-\frac{2}{5}q^2t_c^2  &\frac{1+i\omega t_c}{10} & \frac{2q}{5}(-it_c+t_c^2\omega R+2\omega t_c^2)+\frac{8K}{q}(-t_c^2\omega+it_c-t_c^2\omega R) &0\\
\frac{t_c^2q^2}{5} &-\frac{3-7i}{10}\omega t_c & \frac{it_cq}{5}-\frac{2+R}{5}\omega qt_c^2 &0\\
-it_cq+2\omega t_c^2q  & -\frac{it_cq\omega}{2} & it_c\omega(1+R)-t_c^2\omega^2(1+R+R^2)-\frac{3}{5}q^2t_c^2+24Kt_c^2\\
0 &0 &0 &-\frac{i40Kt_c}{q}
\end{bmatrix}
\begin{bmatrix}
X\\
Y\\
Z\\
W
\end{bmatrix}
=0.
\end{equation}
\end{footnotesize}

Setting the determinant of this matrix to zero and expanding it to first order in $t_c$, we get the dispersion relation as:
\begin{equation}\label{disp}
-5q^2+15(1+R)\omega^2+120K-it_c\omega^3(5+5R-15R^2)+7i\omega t_c q^2-400iK\omega t_c=0
\end{equation}
For solving the above dispersion relation, we split $\omega$ to  real and imaginary part, $\omega=\Omega+i\Gamma$. Insert it in equation (\ref{disp}) and set the real an imaginary part of the relations separately to zero, we get:
\begin{footnotesize}
\begin{equation}\label{realeqn}
-q^2(5+7t_c\Gamma)+400K(3+10t_c\Gamma)-15(1+R)\Gamma^2+5(-1-R+3R^2)t_c\Gamma^3+15(1+R)\Omega^2+15t_c\Gamma\Omega^2(1+R-3R^2)=0
\end{equation}
\end{footnotesize}
\begin{equation}\label{imgeqn}
-400Kt_c+7q^2t_c+30(1+R)\Gamma -15(-1-R+3R^2)t_c\Gamma^2\-5(1+R-3R^2)t_cR^2=0
\end{equation}
Now we can solve (\ref{imgeqn}) for getting $\Omega$:
\begin{equation}\label{omgeqn}
\Omega=\pm \sqrt{\frac{400Kt_c-7q^2t_c+15\Gamma(-2-t_c\Gamma+3R^2t_c\Gamma-R(2+t_c\Gamma))}{5t_c(-1-R+3R^2)}}
\end{equation}
Putting the solution for $\Omega$ back to equation (\ref{realeqn}) and solving it for $\Gamma$ upto first power of $t_c$ we get:
\begin{equation}\label{silk}
\Gamma=-\frac{t_cq^2}{6(1+R)}(\frac{16}{15}+\frac{R^2}{1+R})+\frac{4Kt_c}{1+R}(3+\frac{R^2}{1+R})
\end{equation}
Inserting $\Gamma$ back in equation (\ref{omgeqn}) and expand it to first power of $t_c$ we will get:
\begin{equation}\label{omega}
\Omega=\pm\frac{1}{\sqrt{3(1+R)}} \sqrt{q^2-72K\frac{3+3R+R^2}{-1-R+3R^2}}
\end{equation}
For checking our results, we can look at the limit of $K\rightarrow 0$, where we get back the results for flat universe as it is originally derived in \cite{1983MNRAS.202.1169K}.

\section{Hydrodynamic Solutions in a Spatially Curved Universe}
The 00-component of Einstein equation and conservation equations of energy and momentum are three coupled differential equations which are sufficient for getting hydrodynamic solutions of cosmological perturbations. These equations in synchronous gauge \footnote{Scalar metric perturbation in synchronous gauge is defined like this $ds^2=-dt^2+a^2\left((1+A)\gamma_{ij}+\partial_{ij}B\right) $ } in a curved spacetime are correspondingly:

\begin{equation}\label{00}
-4\pi G(\delta\rho+3\delta p+\nabla^2\Pi^s)=\partial_t(a^2\psi),
\end{equation}

\begin{equation}\label{CE}
\delta p+\nabla^2\Pi^s+\partial_t((\bar{\rho}+\bar{p})\delta u)+3\frac{\dot{a}}{a}(\bar{\rho}+\bar{p})\delta u +2K\Pi^s=0,
\end{equation}

\begin{equation}\label{CM}
\delta\dot{\rho}+3\frac{\dot{a}}{a}(\delta \rho+\delta p)+\nabla^2(\frac{1}{a^2}(\bar{\rho}+\bar{p})\delta u +\frac{\dot{a}}{a}\Pi^s)+(\bar{\rho}+\bar{p})\psi=0
\end{equation}

where $\psi$ is defined as $\psi=\frac{1}{2}(3\dot{A}+\nabla^2\dot{B})$. It is important to note that modification due to spatial curvature only appears here in the form  of $2K\Pi^s$. Ignoring $\Pi^s$, we come to the same equations as we had in flat background. So the known hydrodynamic solutions that was derived in \cite{Weinberg:2008zzc} is applicable here. The fast mode solutions are:
\begin{equation}\label{deltaU}
\delta u_\gamma =\frac{a\sqrt{3}\mathcal{R}_q}{q(1+R)^{3/4}}e^{-\int \Gamma dt}Sin\left(\int \frac{\Omega dt}{a}\right)
\end{equation}

\begin{equation}
\delta_D =48\pi G \bar{\rho}_\gamma(2+R)(1+R)^{3/4}(\frac{a}{q})^2\mathcal{R}^O_q e^{-\int \Gamma dt}Cos
\left(\int \frac{\Omega dt}{a}\right)
\end{equation}
Here,  we should put what we find in (\ref{omega}) and (\ref{silk}) for frequency of oscillation $\Omega$ and damping factor $\Gamma$. Ignoring photon and neutrino energy density combining equations (\ref{00}),(\ref{CE}) and (\ref{CM}) we come to a second order differential equation:

\begin{equation}
\frac{d}{dt}(a^2\frac{d}{dt}{\delta_M})=4\pi G a^2\bar{\rho}_M{\delta_M}
\end{equation}
We can factorize the dependence of t and q by writing: $\delta _M=\Delta(q) F(t)$, where $\Delta$ satisfies:
\begin{equation}\label{Delta}
\Delta(q)=\beta {\delta_\gamma}_q(t_L)+(1-\beta){\delta_D}_q(t_L)-t_L\psi_q(t_L)+\beta t_L \frac{q^2}{a^2}\delta {u_\gamma}_q(t_L)
\end{equation}

 So the time evaluation  of $F(t)$ is:
 \begin{equation}\label{eq1}
 \frac{d}{dt}(a^2\frac{dF}{dt})=4\pi G a^2 \bar{\rho}_M F
 \end{equation}
with initial condition $F \rightarrow \frac{3}{5}(\frac{t}{t_L})^{2/3}$.
From Friedman equation we have:
\begin{equation}
(\frac{\dot{a}}{a})^2=\frac{8\pi G}{3}(\rho_\Lambda+\bar{\rho}_M)-\frac{K}{a^2}
\end{equation}
Defining $X\equiv \frac{\rho_\Lambda}{\bar{\rho}_M}=\frac{\Omega_\Lambda}{\Omega_M}(\frac{a}{a_0})^3$, The Friedman equation can be written in terms of $X$:
\begin{equation}\label{eq2}
\frac{\dot{a}}{a}=H_0\sqrt{\Omega_\Lambda}\sqrt{1+\frac{\Omega_K}{\Omega_\Lambda^{1/3}\Omega_M^{2/3}}\frac{1}{X^{2/3}}+\frac{1}{X}}
\end{equation}
With the help of (\ref{eq2}), the differential equation (\ref{eq1}) can be written in terms of X. Then the solutions as function of X is:

\begin{equation}
F\propto \sqrt{\frac{1+\frac{\Omega_K}{\Omega_\Lambda^{1/3}\Omega_M^{2/3}}X^{1/3}+X}{X}}\int_0^x\frac{du}{u^{1/6}(1+\frac{\Omega_K}{\Omega_\Lambda^{1/3}\Omega_M^{2/3}}u^{1/3}+u)^{3/2}}
\end{equation}

We know that $F\rightarrow \frac{3}{5}\frac{a}{a_L}$ at early times. In this way, we can set coefficient of proportionality and write: 
\begin{equation}
F=\frac{3}{5}\frac{a(t)}{a_L}C(\frac{\Omega_\Lambda}{\Omega_M}(\frac{a}{a_0})^3)
\end{equation}
where,
\begin{equation}
C(X)\equiv \frac{5}{6}X^{-5/6}\sqrt{1+\frac{\Omega_K}{\Omega_\Lambda^{1/3}\Omega_M^{2/3}}X^{1/3}+X}\int_0^x\frac{du}{u^{1/6}(1+\frac{\Omega_K}{\Omega_\Lambda^{1/3}\Omega_M^{2/3}}u^{1/3}+u)^{3/2}}
\end{equation}

We can calculate this integral numerically. The result for different values of $X$ is written in the Table \ref{tab:C(x)}.
As it can be seen from the table, for the flat universe, $C(X)$ just have a suppression effect on growth of matter due to dark energy. In the closed case, for small values of $X$, $C(X)$ gives enhancement and then it gives suppression.\\
\begin{table}
\begin{center}
 \begin{tabular}{|c | c| c|} 
 \hline
 X & C(X) Closed & C(x) Flat \\ [0.5ex] 
 \hline\hline
 0.1 & 1.0113 & 0.9826 \\ 
 \hline
 0.2 & 1.0013 & 0.9667 \\  
 \hline
 0.3 & 0.9901 & 0.9520\\
 \hline
 0.5 & 0.9675 & 0.9256\\
 \hline
 0.7 & 0.9463 & 0.9025\\
 \hline
 1.0 & 0.9176 & 0.8725\\
 \hline
 1.5 & 0.8769 & 0.8314\\
 \hline
 2.0 & 0.8432 & 0.7981\\
 \hline
 2.5 & 0.8146 & 0.7702\\
 \hline
 3.0 & 0.7899 & 0.7462\\
 \hline
 3.5 & 0.7683 & 0.7254\\[1ex] 
 \hline
\end{tabular}
\end{center}
\caption{\label{tab:C(x)}The values of C(x) which is the effect  of dark energy on  growth of matter, as a function of $X \equiv \frac{\Omega \Lambda}{\Omega_M}$ for closed and flat spacetime, We use 2018 Planck data \cite{Aghanim:2018eyx} for its calculation, which is $\Omega_M=0.315$ and for closed case $\Omega_K=-0.044$.}
\end{table}
Furthermore ignoring baryon's effects (e.g. neglecting the terms of order $\beta$), from(\ref{Delta}) we have $\Delta(q)={\delta_D}_q(t_L)-t_L\psi_q(t_L)$, then we would get  the solution in the form of $
\Delta(q)=\frac{2q_n^2{\cal R}^O_n\tau(\kappa)}{3H_L^2a_L^2}$, the same as flat universe (where $\tau$ is transfer function, $t_L=\frac{2}{3H_L}$ and $H_L=\sqrt{\Omega_M}H_0(1+z_L)^{3/2}$).  This will give us well-known hill-top shape power spectrum. When we consider the effects of the baryons, then what is the dominant term in (\ref{Delta}) for fast mode is $\beta t_L \frac{q_n^2}{a^2}\delta u_\gamma(t_L)$, so using the fast mode solution of (\ref{deltaU}),we can write the q dependent part of matter density as:
\begin{footnotesize}
\begin{equation}
\Delta(q)\approx \beta t_L (\frac{q}{a_L})^2 \delta u_\gamma=\frac{2\beta q \mathcal{R}_q^O}{\sqrt{3}a_LH_L(1+R_L)^{3/4}}e^{-\int \Gamma dt} Sin\left(\int \frac{q dt}{a\sqrt{3(1+R)}}\left(1-\frac{72K}{q^2}\frac{3+3R+R^2}{-1-R-3R^3} \right)\right)
\end{equation}
\end{footnotesize}
Here also we should use the modified form of $\Gamma$ as (\ref{silk}). This is the well-known acoustic oscillation forced by baryon's effect. 

\section{Conclusion}
\label{sec:con}
In this work, first we write, Boltzmann equation for dark matter particles in a spatially curved spacetime. The integral solution to this equation shows that memory of gravitational field at early time affects growth at later times. Specifically it is shown that  spatial curvature modified this memory by the factor that is come in the last term of (\ref{end1}). Then we write Boltzmann equation for photons and using it find a dispersion relation in the presence of spatial curvature. This gives us damping factor and frequency of acoustic oscillations, as (\ref{silk}) and (\ref{omega}). We want to emphasis that although there is extensive literature on the subject of Boltzmann equation in non-flat FRW universe, but the effect of spatial curvature on the mentioned phenomena was not explored specifically.  In the last section we show that how results of section 3, affect evolution of matter density in the hydrodynamic regime. As it is shown, in this hydrodynamic regime  spatial curvature does not affect $q$-dependent part of matter density, $\Delta(q)$. But for the time dependent part, in spite of the fact that dark energy always gives a factor of suppression to the growth of matter in a flat universe, but here when ratio of dark energy to matter is small it gives an enhancement and only when this ratio becomes larger, it changes to suppression.  When considering baryons, effects of spatial curvature shows itself in terms of modification of frequency and damping parameters (the results of section 3) of acoustic oscillations.

\appendix
\section{Further Computations}\label{app:slow}

The useful identities for following calculation is:
\begin{equation}
\int d^2\hat{p} =4\pi
\end{equation}
\begin{equation}
\int d^2\hat{p} \hat{p}_i\hat{p}_j=\frac{4\pi}{3}\delta_{ij}
\end{equation}
\begin{equation}
\int d^2\hat{p} \hat{p}_i\hat{p}_j\hat{p}_k\hat{p}_l=\frac{4\pi}{15}(\delta_{ij}\delta_{kl}+\delta_{ik}\delta_{jl}+\delta_{il}\delta_{kj})
\end{equation}

Here we write down the non-zero results of integration of each term in (\ref{boltint}):

\begin{equation}
\int d^2\hat{p}(\delta_{ij}-\hat{p}_i\hat{p}j)X=4\pi X(\delta_{ij}-\frac{1}{3}\delta_{ij})=\frac{8\pi}{3}X\delta_{ij}
\end{equation}

\begin{equation}
it_c\int d^2\hat{p}(\delta_{ij}-\hat{p}_i\hat{p}j)\mathcal{O}X=it_c\omega X\delta_{ij}\int d^2\hat{p}-it_cXw\int d^2\hat{p}\hat{p}_i\hat{p}_j=\frac{8\pi i t_c Xw}{3}\delta_{ij}
\end{equation}

\begin{equation}
\begin{split}
-t_c^2\int d^2\hat{p}(\delta_{ij}-\hat{p}_i\hat{p}_j)\mathcal{O}^2X&=-t_c^2\int d^2\hat{p}(\delta_{ij}-\hat{p}_i\hat{p}_j)(\omega^2X-2\omega\hat{q}_i\hat{p}_jX+\hat{q}_i\hat{q}_j\hat{p}_i\hat{p}_jX-8KX)\\
&=-t_c^2\frac{8\pi}{3}(\omega^2X-8KX)\delta_{ij}-\frac{16\pi}{15}t_c^2q^2X\delta_{ij}+\frac{8\pi}{15}t_c^2Xq^2\hat{q}_i\hat{q}_j
\end{split}
\end{equation}

\begin{footnotesize}
\begin{equation}
\begin{split}
it_c Z(1+it_c\omega R)\int d^2\hat{p}(\delta_{ij}-\hat{p}_i\hat{p}_j)\mathcal{O}\hat{p}.\hat{q}&=it_cZ(1+i\omega t_c R)\int d^2\hat{p}(\delta_{ij}-\hat{p}_i\hat{p}_j)(\omega \hat{p}.\vec{q}-\hat{p}.\vec{q}\hat{p}.\vec{q}+\frac{8K}{q})\\
&=it_cZ(1+i\omega t_c R)(\frac{8K}{q}\frac{8\pi}{3}\delta_{ij}-\frac{16\pi}{15}q^2\delta_{ij}+\frac{8\pi}{15}q^2\hat{q}_i\hat{q}_j)
\end{split}
\end{equation}
\end{footnotesize}

\begin{equation}
\begin{split}
-t_c^2\int d^2\hat{p}(\delta_{ij}-\hat{p}_i\hat{p}_j)Z\mathcal{O}^2\hat{p}.\hat{q}&=-t_c^2Z \int d^2\hat{p}(\delta_{ij}-\hat{p}_i\hat{p}_j)(-2\omega q_k\hat{q}_{k'}\hat{p}_k\hat{p}_{k'}+\frac{8K\omega}{q})\\
&=-t_c^2Z(-2\omega\frac{4\pi}{3}q^2\delta_{ij}+2\omega \frac{4\pi}{15}q^2(\delta_{ij}+2\hat{q}_i\hat{q}_j)+\frac{8K\omega}{q}\frac{8\pi}{3}\delta_{ij})\\
&=-t_c^2Z(\frac{32\pi}{15}\omega q^2\delta_{ij}+\frac{64\pi K\omega}{3q}\delta_{ij}+\frac{16\pi}{15}\omega\hat{q}_i\hat{q}_j)
\end{split}
\end{equation}

\begin{equation}
Y\int d^2\hat{p}(\hat{q}^i-\hat{p}^i \hat{p}_k\hat{q}_k)(\hat{q}^j-\hat{p}^j\hat{p}_{k'}\hat{q}_{k'})=\frac{4\pi}{15}Y(\delta_{ij}+7\hat{q}_i\hat{q}_j)
\end{equation}

\begin{equation}
it_cY\int d^2\hat{p}\mathcal{O}(\hat{q}^i-\hat{p}^i \hat{p}_k\hat{q}_k)(\hat{q}^j-\hat{p}^j\hat{p}_{k'}\hat{q}_{k'})=it_cY(\frac{28\pi}{15}\omega \hat{q}_i\hat{q}_j+\frac{4\pi}{15}\omega\delta_{ij})
\end{equation}

The calculation of integration for each term in (\ref{forapp}) is presented here:

\begin{equation}
(Z+i\omega t_c R Z(1+i\omega t_c R)-\frac{it_c40KW}{q})\hat{q}_i\int d^2\hat{p}\hat{p}_i\hat{p}_j=\frac{4\pi}{3}((1+i\omega t_cR-\omega^2 t_c^2R^2)Z-\frac{it_c40KW}{q})\hat{q}_j
\end{equation}

\begin{equation}
it_c\int d^2\hat{p}\hat{p}_j\mathcal{O}(X+\frac{Y}{2})=-\frac{4\pi}{3}it_c(X+\frac{Y}{2})\hat{q}_j
\end{equation}

\begin{footnotesize}
\begin{equation}
\begin{split}
-t_c^2\int d^2\hat{p}\hat{p}_j\mathcal{O}^2(X+\frac{Y}{2})=-t_c^2\int d^2\hat{p}\hat{p}_j \Bigl((X+\frac{Y}{2})\omega^2-2\omega(X+\frac{Y}{2})\vec{q}.\hat{p}+(X+\frac{Y}{2})&\vec{q}.\hat{p}\vec{q}.\hat{p}-8K(X+\frac{Y}{2})\Bigr)\\
&=\frac{8\pi}{3}\omega t_c^2(X+\frac{Y}{2})q\hat{q}_j
\end{split}
\end{equation}
\end{footnotesize}

\begin{footnotesize}
\begin{equation}
\begin{split}
it_c\int d^2\hat{p}\hat{p}_j\mathcal{O}(Z+it_c\omega RZ-\frac{Y}{2})\hat{p}.\hat{q}&=it_c\int d^2\hat{p}\hat{p}_j\Bigl(\omega \hat{q}.\hat{p}(Z-\frac{Y}{2}+it_c\omega RZ)-\vec{q}.\hat{p}\hat{q}.\hat{p}(Z-\frac{Y}{2}+i\omega R t_c Z)\\
&+\frac{8K}{q}(Z-\frac{Y}{2}+i\omega R t_cZ) \Bigr)
=it_c\omega \frac{4\pi}{3}(Z-\frac{Y}{2}+it_c\omega RZ)\hat{q}_j
\end{split}
\end{equation}
\end{footnotesize}

\begin{equation}
-t_c^2\int d^2\hat{p} \hat{p}_j\mathcal{O}^2Z\hat{p}.\hat{q}=-4\pi\Bigl(\frac{t_c^2}{3}(\omega^2q-8K(1+q))+\frac{t_c^2}{5}q^3 \Bigr)Z
\end{equation}


\end{document}